\documentclass[twocolumn]{revtex4}
\usepackage{graphicx, amssymb}

\begin{document}

\title{Standing shocks in the inner slow solar wind}

\thanks{This research is supported by grant NNSFC 40904047,
     and also by the Specialized Research Fund for State Key Laboratories.
     $^{**}$Email: bbl@sdu.edu.cn}

\author{Bo Li$^{1,2}$, Yan-Jun Chen$^{1}$, Xing Li$^{3}$ \\
{\small $^{1}$ Shandong Provincial Key Laboratory of Optical Astronomy \& Solar-Terrestrial Environment,} \\
{\small  School of Space Science and Physics, Shandong University at Weihai, Weihai 264209, China}  \\
{\small $^{2}$ State Key Laboratory of Space Weather, Chinese Academy of Sciences, Beijing 100190, China} \\
{\small $^{3}$ Institute of Mathematics and Physics, Aberystwyth University, Aberystwyth SY23 3BZ, UK}
}

\date{Received: 14 Dec 2010, Accepted: 24 Mar 2011}
\begin{abstract}
We examine whether the flow tube along the edge of a coronal streamer
supports standing shocks in the inner slow wind by solving an isothermal wind
model in terms of the Lambert W function.
We show that solutions with standing shocks do exist, and they exist in a broad area in the parameter space
characterizing the wind temperature and flow tube. In particular,
streamers with cusps located at a heliocentric distance $\gtrsim 3.2
R_\odot$ can readily support discontinuous slow winds with
temperatures barely higher than $1$\,MK.\\

\end{abstract}

\maketitle

It was proven possible that the
quasi-steady solar wind may not be continuous but involve standing
shocks in the near-Sun region\cite{Holzer_77,HasanVenka_82,
HabbalTsing_83,HabbalRosner_84,Habbal_etal_94,LeerHolzer_90,MarschTu_97}!
First pointed out 30 years ago\cite{Holzer_77},
the existence of standing shocks depends critically
on the existence of multiple critical points (CPs).
These can arise due to either momentum addition or rapid tube
expansion near the base.
Time-dependent simulations showed that 
whether the system adopts a continuous or a discontinuous solution depends on the detailed manner the tube
geometry is varied\cite{HasanVenka_82,Habbal_etal_94}, or how the
momentum addition is applied\cite{HabbalRosner_84,Habbal_etal_94}.
Existing studies on standing shocks were exclusively on the
flow rooted in the interior of coronal holes. However, little is known about whether the flow
 tubes bordering bright streamer helmets can support
standing shocks as well. This region is important, however, since it is 
where the slow wind likely originates\cite{YMWang_etal_09}. Here the 
tube expansion is distinct from the coronal-hole one, with  
the tube likely to experience a dramatic expansion around the streamer
cusp (see Fig.4, the current-sheet case in\cite{WangSheeley_90}).  
This letter is intended to answer: Are standing shocks allowed by this geometry? 

To isolate the geometrical effect, we will use a simple isothermal model.
Let $T$ and $v_r$ denote the solar wind temperature and radial speed, respectively. 
The isothermal sound speed is then $c_s=\sqrt{2k_B
T/m_p}$, where $k_B$ is the Boltzmann constant,
 and $m_p$ the proton mass.
The Mach number $M=v_r/c_s$ is governed by\cite{Velli_01}
\begin{eqnarray}
   \left(M-\frac{1}{M}\right)\frac{d M}{d y}=
   \frac{d \ln \bar{a}}{d y}-\frac{\Delta}{y^2},
   \label{eq_v_nondim}
\end{eqnarray}
where $y=r/R_\odot$, with $R_\odot$ the solar radius and $r$ the
heliocentric distance. Moreover, $\bar{a}=a/R_\odot^2$ is the non-dimensionalized
tube cross-section $a$. And $a$ is related to the expansion factor $f$ by
 $a(r) = f(r) r^2$.
Furthermore, $\Delta = g_\odot R_\odot/c_s^2$ where $g_\odot$ is the
surface gravitational acceleration. Evidently $\Delta$
measures the relative importance of the gravitational force and
pressure gradient force.

\begin{figure}[htp]
\centering
\includegraphics[width=0.95\columnwidth]{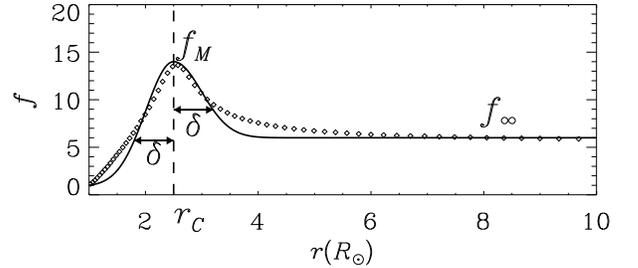}
\caption{Expansion factor $f$ for the streamer geometry vs.
   heliocentric distance $r$.
Please see text for the meaning of $f_\infty$, $f_M$, $r_C$, and $\delta$,
   and what the diamonds refer to.}
\label{fig_geometry}
\end{figure}

The streamer geometry is parameterized as
\begin{eqnarray}
f(r) =  \left\{ \begin{array}{l}
1 + \left(f_M-1\right) \frac{G(r; r_C, \delta) -G(R_\odot; r_C, \delta) }{1-G(R_\odot; r_C, \delta)},
              \\
    \hspace{0.5cm} r \le r_C ,   \\
f_\infty + \left(f_M- f_\infty\right) G(r; r_C, \delta),
           \\
\hspace{0.5cm}  r \ge r_C ,
     \end{array}\right.
\end{eqnarray}
where $G(x;x_0, \delta) = \exp\left[-\left(x-x_0\right)^2/\delta^2
\right]$ is a Gaussian. Figure~\ref{fig_geometry} illustrates the
$r$-distribution of $f$. Obviously $f_{\infty}$ represents the
value at large distances, and $f_M$ is the maximum
attained at $r_C$, the heliocentric distance of the streamer cusp.
Moreover, $\delta$ describes how rapid $f_M$ is approached. For $r_C$, we adopt values between
$2.4$ and $3.6$$R_\odot$, compatible with LASCO C2 images. The
ranges for $f_\infty$, $f_M$ and $\delta$ are $[2,
10]$, $[6, 22]$, and $[0.4, 1]$$R_\odot$, respectively. 
As direct measurements of the coronal magnetic field remain largely
unavailable, some model field is used to guide our choice.
The $f$ profile with the base values ($f_\infty=6, f_M=14$, and
$\delta=0.7$$R_\odot$)
is close to the diamonds in Fig.\ref{fig_geometry}, which
correspond to $f$ along the tube at the streamer edge
in a current sheet model, given
in Fig.4b of\cite{WangSheeley_90} (the one labeled 27$^\circ$).

Given the temperature $T$ and an $f(r)$, the
right hand side (RHS) of Eq.(\ref{eq_v_nondim}) can be readily
evaluated and determines whether solutions with standing shocks are
allowed. To explain this, we note that any root of
$\mbox{RHS}=0$ corresponds to a critical point (CP), which is either
a local extreme ($dM/dy=0, M\ne 1$) or a sonic point (SP) ($dM/dy
\ne 0, M=1$, denoted by the subscript $S$). Shock solutions are 
known to exist only when there are multiple CPs, and were usually
constructed by carefully examining the solution topology. Here we
present a new method based on a recent study which shows that a
transonic solution to Eq.(\ref{eq_v_nondim}) is expressible in terms
of the Lambert W function $W(x)$\cite{Cranmer_04}
  \begin{eqnarray}
   M^2 =  \left \{ \matrix{
    & -W_{0}(-D(y)) , & 1 \le y \le y_S, \cr
    & -W_{-1}(-D(y)) , &  y \ge y_S,   \cr
      }\right.
   \label{eq_solution_in_LW}
   \end{eqnarray}
where
  \begin{eqnarray}
  D\left(y\right)=\frac{\bar{a}_S^2}{\bar{a}^2}
  \exp\left[2\Delta\left(\frac{1}{y_S}-\frac{1}{y}\right)-1\right].
  \label{eq_def_D}
  \end{eqnarray}
Only two things about $W(x)$ need to be known in the present
context:
 first, a real-valued $W(x)$ can be
 defined only for $x \ge -1/e$ (note that $D$ is positive definite);
 second, $W(x)$ has two branches for $-1/e < x <0$,
 and they obey $-1\le W_0 <0$ and $W_{-1} \le -1$.
The mathematical details can be found in\cite{Corless_etal_96}. 
In practice, we evaluate $W_0$ and $W_{-1}$ via Eq.(5.9) there.

If only one CP exists, it is naturally the SP, and
Eq.(\ref{eq_solution_in_LW}) describes the only possible transonic
solution for which $M$ increases monotonically with $r$. This is the
case considered in\cite{Cranmer_04}, where $f\equiv 1$ is assumed.
In our case there exist up to 3
CPs, and hence we have to extend the Lambert W function approach
as follows. First, when 3 CPs exist, only the innermost and
outermost ones turn out relevant. We evaluate $D$ by choosing each
of them, one after another, as the SP. In some portion of the
computational domain ($y\ge 1$), $D$ for one CP may exceed $1/e$,
and hence the solution is not defined.
Call this solution the ``broken solution'', denoted by $M_b$.
Choosing the other CP as SP results in a continuous solution, denoted by $M_c$.
If standing shocks exist, they have to
appear where the Rankine-Hugoniot relations and evolutionary
condition are met. In the isothermal case these translate
into\cite{HabbalRosner_84,Velli_01}
\begin{eqnarray}
M^{+} M^{-} = 1  \mbox{ and } M^{+} >1 ,
\label{eq_shock_JC}
\end{eqnarray}
respectively. Here $+$ ($-$) represents the shock upstream
(downstream). This suggests a simple graphical means to construct
solutions with shocks\cite{Velli_01}, where we plot $M_b$, and
examine whether it intersects the $1/M_c$ curve. Any intersection
represents a shock jump, however the solution cannot jump from a
lower to a higher curve.

\begin{figure}[htp]
\centering
 \includegraphics[width=0.95\columnwidth]{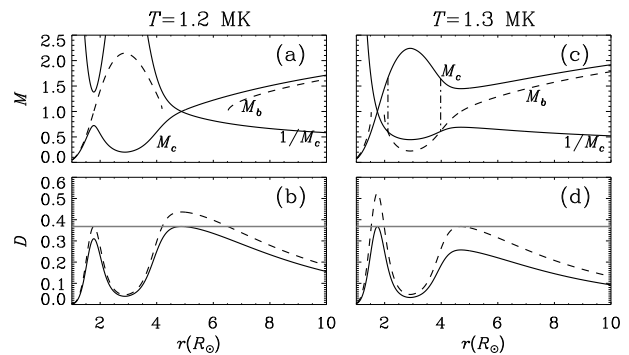}
\caption{Wind solutions with the streamer geometry for which
$f_\infty=6$, $f_M=14$, $r_C = 3$$R_\odot$,
and $\delta = 0.7$$R_\odot$.
In the left and right columns, $T$ is 1.2 and 1.3\,MK, respectively.
Panels (a) and (c) present the $r-$dependence of the Mach number $M$, while
(b) and (d) give that of $D$.
The solid (dashed) lines represent the continuous (broken) solutions.
In (b) and (d), the horizontal lines give $1/e$ for comparison.
Standing shocks are allowed only when the two curves $1/M_c$ and $M_b$ intersect.
}
\label{fig_shock_or_not}
\end{figure}

Figure~\ref{fig_shock_or_not} illustrates our solution procedure,
giving the radial dependence of the Mach number $M$ ((a) and (c))
and $D$ ((b) and (d)). In Figs.\ref{fig_shock_or_not}b and
\ref{fig_shock_or_not}d, the light horizontal lines represent $1/e$.
The solid and dashed lines correspond to the continuous and broken
solutions, respectively. In addition to $M$,
Figs.\ref{fig_shock_or_not}a and \ref{fig_shock_or_not}c also give
$1/M_c$. Figures~\ref{fig_shock_or_not}a and \ref{fig_shock_or_not}b
are for $T=1.2$\,MK, while Figs.\ref{fig_shock_or_not}c and
\ref{fig_shock_or_not}d are for $T=1.3$\,MK. In both cases the
tube parameters are $f_\infty=6$, $f_M=14$, $r_C=3$$R_\odot$,
and $\delta = 0.7$$R_\odot$. Consider now
Figs.\ref{fig_shock_or_not}a and \ref{fig_shock_or_not}b. It is seen
 that both curves in {Fig.\ref{fig_shock_or_not}b}
exhibit three local extrema, whose locations correspond to the CPs.
This follows from that $d D/d y=0$ at any CP (see
Eq.(\ref{eq_def_D})). Furthermore, the global maximum of $D$ is
attained at the outermost CP, located at 4.89$R_\odot$. Therefore
when the innermost CP is chosen as the SP, $D>1/e$ around the
outermost CP for $4.2 \le r \le 6.51$$R_\odot$. Recalling that
$W(-D)$ is real-valued only when $-D \ge -1/e$, one readily
understands that in this interval choosing the innermost CP as the SP
does not result in a solution to Eq.(\ref{eq_v_nondim}).
Figure~\ref{fig_shock_or_not}a also shows that the curve $1/M_c$ does not intersect $M_b$, indicating
 the solution to Eq.(\ref{eq_v_nondim}) is unique and is the continuous one.

The situation changes when $T=1.3$\,MK. Now the global maximum of
$D$ is attained at the innermost CP, located at 1.75$R_\odot$
(Fig.\ref{fig_shock_or_not}d). Choosing the outermost CP as the SP
leads to that $D>1/e$ in the interval [1.53, 1.98]$R_\odot$ where
there is no solution (Fig.\ref{fig_shock_or_not}c). However,
two standing shocks are now allowed, since two crossings exist
between the curves $1/M_c$ and $M_b$, located at 2.11 and
3.96$R_\odot$, respectively. Hence in addition to the continuous one
($M_c$ adopting the innermost CP as the SP), two
additional solutions exist to Eq.(\ref{eq_v_nondim}): both start with
$M_c$ but one connects to $M_b$ at the inner crossing, the other
connects to $M_b$ at the outer one.

\begin{figure}[htp]
\centering
 \includegraphics[width=0.8\columnwidth]{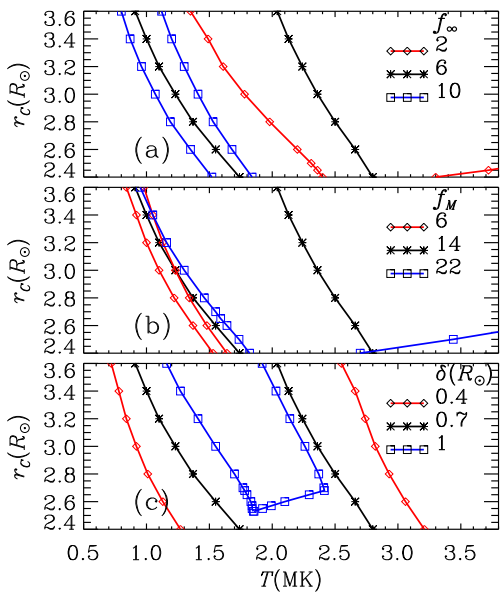}
\caption{Regions in the $T-r_C$ space where standing shocks are allowed, shown as
the area bounded by the two curves in same color.
The results are obtained by varying $f_\infty$, $f_M$ and $\delta$ respectively,
 the rest of the parameters are fixed at their reference values
 $f_\infty = 6$, $f_M=14$, $\delta=0.7$$R_\odot$.
}
\label{fig_para_shock_or_not}
\end{figure}

Although Eq.(\ref{eq_v_nondim}) permits solutions with shocks,
and time-dependent simulations suggest these steady-state
solutions can be attained\cite{HasanVenka_82,
HabbalRosner_84,Habbal_etal_94,MarschTu_97}, one may
still question whether the shock solutions can stand the
sensitivity test similar to\cite{LeerHolzer_90} which showed
standing shocks in the solar wind from the center of coronal holes are very unlikely,
 for the parameter range allowing shock solutions is extremely limited. To see whether the
same happens with the streamer geometry, we note that given
$f_\infty$, $f_M$ and $\delta$, shock solutions are
allowed only in the area bounded by two curves in the $[T, r_C]$ space.
Figure~\ref{fig_para_shock_or_not} presents a series of such curves
obtained by varying (a) $f_\infty$, (b) $f_M$, and (c) $\delta$
about the reference values $f_\infty = 6$, $f_M=14$, and
$\delta=0.7$$R_\odot$. (In what follows, the temperatures are in MK,
$r_C$ and $\delta$ in~$R_\odot$.) Let us first examine the cases with
reference values (the black curves connecting asterisks).
Figure~\ref{fig_para_shock_or_not} shows that the area bounded by
the two curves is rather broad, and with increasing $r_C$, both
curves are shifted towards lower temperatures, indicating streamers
whose cusps are located higher in the corona are more likely
associated with standing shocks. For instance, 
when the cusp is located at $3.6$$R_\odot$, the slow wind may
possess standing shocks as long as $0.91 \le T \le 2.04$, which actually
 tends to be lower than the often-quoted values of coronal temperatures. 
On the other hand, even for the lowest cusp height examined ($r_C=2.4$$R_\odot$), the temperature
range is $[1.74, 2.80]$, still largely compatible with
the observational range.

Figure~\ref{fig_para_shock_or_not}a examines the effects of varying
$f_\infty$.
It is seen that while $f_\infty$ decreases from its
reference value to $2$ (the red curves), the $T$ range allowing
shocks increases significantly. Actually for $r_C \ge 2.5$,
in the examined temperature range there virtually exists no upper
bound for shocks to occur. Take $r_C=3.6$ for instance.
Shock solutions take place as long as $T\ge 1.35$. On the contrary,
increasing $f_\infty$ to $10$ (the blue curves) makes shocks
appear in a much narrower $T$ range (the width is $\sim 0.3$).
The effects of varying $f_M$ are shown in
Fig.\ref{fig_para_shock_or_not}b, which shows that increasing $f_M$
considerably broadens the area allowing standing shocks. For
example, with $f_M$ increasing from $6$ to $14$, the width along the
$T$-axis of the area increases from $\sim 0.1$ to $1.1$. When $f_M$
further increases to $22$, this width increases dramatically from
$\sim 0.88$ at $r_c = 2.4$ to $\gtrsim 2.9$ for $r_c \ge 2.6$.
Figure~\ref{fig_para_shock_or_not}c shows what happens when $\delta$
changes, where it is seen that increasing $\delta$ reduces the
range of $T$ where shocks are allowed. For instance, with
$r_C=3.0$, this range for $\delta=0.4$ ($\delta=1$) is $[1.27,
3.21]$ ($[1.55, 2.26]$), while the range for the reference value
$\delta=0.7$ lies in between. It is interesting to note that
for $\delta=1$, at $r_C \sim 2.68$ the upper bound for $T$ (the
right blue curve) changes its slope dramatically, and for $r_C \le
2.53$ no shock solutions exist. For $2.53 \le r_C
\le 2.68$, it turns out that on the right of the right blue curve
actually no solution exists, since now only two critical
points exist and neither of them corresponds to a $D\le 1/e$
throughout the computational domain (see
Eq.(\ref{eq_solution_in_LW})). This is different from the portion
$r_C \ge 2.68$, where on the right of the right blue curve there
does exist a solution which is the continuous one. Putting the three
panels together, one may see that for most combinations of tube
parameters, the area in the $T-r_C$ space supporting standing
shocks is substantial. Hence with the streamer
geometry, standing shocks in the inner slow wind seem physically
accessible.

It is not easy to exhaust the possible tube parameters and the
consequent changes in shock properties. Let us instead discuss
only the shocks found, examining their detectability. First, $\delta \rho$, the density jump relative to the upstream value, is up 
to $8$, a result of the isothermal assumption
exceeding the nominal upper limit of $4$ for adiabatic gases.
As shown by\cite{EsserHabbal_90}, a $\delta \rho$ of $\sim 2.3$ at a
standing shock produces an enhancement in the polarized brightness
intensity that is only marginally detectable. A $\delta \rho$ of $8$
certainly makes such detections easier, but one can not say this for
sure without constructing detailed observables. Second, by conserving angular 
momentum a coronal shock also produces a discontinuity in the azimuthal flow speed $v_\phi$,
leading in principle to measurable Doppler shifts in H I Ly $\alpha$.
However, the jump in $v_\phi$ turns out $\lesssim 4$~km/s,
discerning which is way beyond the sensitivity of SOHO/UVCS, 
whose spectral resolution of $0.23$~\AA\ translates into $\sim 57$~km/s.

The isothermal assumption needs some justification.
First, it is not far from reality. The UVCS measurements of the H I
Ly $\alpha$ emission from an equatorial
streamer\cite{Strachan_etal_02} showed that the proton kinetic
temperature $T_p$ in the stalk decreases only mildly from
1.45\,MK at 3.6$R_\odot$ to 1.3\,MK at 5.1$R_\odot$ (their Fig.3b).
If the stalk and one of streamer legs are on
the same flow tube, then Fig.4b in\cite{Strachan_etal_02} shows that
$1.41 \le T_p \le 2.09$\,MK at 2.33$R_\odot$ (the leftmost
two open circles and rightmost two solid ones in their Fig.4d).
As for $T_e$, the electron-scattered
H I Ly$\alpha$ measured by UVCS yielded a
$T_e$ of $1.1 \pm 0.3$\,MK at 2.7$R_\odot$\cite{Fineschi_etal_98}. Although for a streamer,
this value may serve to estimate $T_e$ in flowing regions at
similar heights. Direct $T_e$ measurements 
above that distance are sparse. Nonetheless, multi-fluid MHD
models indicate that $T_e$ ranges from 0.8\,MK at 3$R_\odot$ to
0.65\,MK at 5$R_\odot$ (Fig.3d in\cite{Li_etal_06}). The mean of
$T_e$ and $T_p$, the temperature $T$ in this study is thus $\sim
1.1-1.8$\,MK at 2.3$R_\odot$ and decreases to $\sim
1$\,MK at $5$$R_\odot$. Furthermore, $T$ at the slow wind source
region is $\sim 0.8-1.2$\,MK, be this source in a coronal
hole or in its neighboring quiet Sun\cite{Habbal_etal_93}. 
Second, introducing a more complete energy equation, as was done
in\cite{Habbal_etal_94} for a coronal-hole flow, will likely {\it
strengthen} rather than weaken our conclusion. That study shows that
introducing thermal conduction and two-fluid
effects allows for a much broader parameter range supporting standing
shocks, compared with isothermal and polytropic
computations.


\begin{thebibliography}{0}
\bibitem{Holzer_77}
Holzer T~E 1977
J. Geophys. Res. {\bf 82} 23 %DOI: 10.1029/JA082i001p00023

\bibitem{HasanVenka_82}
Hasan S~S and Venkatakrishnan P 1982
Sol. Phys. {\bf 80} 385 %DOI: 10.1007/BF00147985

\bibitem{HabbalTsing_83}
Habbal S~R and Tsinganos K 1983
J. Geophys. Res. {\bf 88} 1965 %DOI: 10.1029/JA088iA03p01965

\bibitem{HabbalRosner_84}
Habbal S~R and Rosner R 1984
J. Geophys. Res. {\bf 89} 10645 %DOI: 10.1029/JA089iA12p10645

\bibitem{Habbal_etal_94}
Habbal S~R, Hu Y~Q and Esser R 1994
J. Geophys. Res. {\bf 99} 8465 %DOI: 10.1029/94JA00349

\bibitem{LeerHolzer_90}
Leer E and Holzer T~E 1990
Astrophys. J. {\bf 358} 680 %DOI: 10.1086/169021

\bibitem{MarschTu_97}
Marsch E and Tu C-Y 1997
Sol. Phys. {\bf 176} 87 %DOI: 10.1023/A:1004975703854

\bibitem{YMWang_etal_09}
Wang Y-M, Ko Y-K and Grappin R 2009
Astrophys. J. {\bf 691} 760 %DOI: 10.1088/0004-637X/691/1/760

\bibitem{WangSheeley_90}
Wang Y-M and Sheeley N R Jr. 1990
Astrophys. J. {\bf 355} 726 %DOI: 10.1086/168805

\bibitem{Cranmer_04}
Cranmer S~R 2004
Am. J. Phys. {\bf 72} 1397 %DOI: 10.1119/1.1775242

\bibitem{Corless_etal_96}
Corless R~M, Gonnet G~H, Hare D~E~G, Jeffrey D~J and Knuth D~E 1996
Adv. Comput. Math. {\bf 5} 329 %DOI: 10.1007/BF02124750

\bibitem{Velli_01}
Velli M 2001
Astrophys. Space Sci. {\bf 277} 157 %DOI: 10.1023/A:1012237708634

\bibitem{EsserHabbal_90}
Esser R and Habbal S~R 1990
Sol. Phys. {\bf 129} 153 %DOI: 10.1007/BF00154371

\bibitem{Strachan_etal_02}
Strachan L, Suleiman R, Panasyuk A~V, Biesecker D~A and Kohl J~L 2002
Astrophys. J. {\bf 571} 1008 %DOI:10.1086/339984

\bibitem{Fineschi_etal_98}
Fineschi S , Gardner L~D, Kohl J~L, Romoli M and Noci G 1998
Proc. SPIE {\bf 3443}  67 %DOI: 10.1117/12.333614

\bibitem{Habbal_etal_93}
Habbal S~R, Esser R and Arndt M~B 1993
Astrophys. J. {\bf 413} 435 %DOI: 10.1086/173011

\bibitem{Li_etal_06}
Li B, Li X and Labrosse N 2006
J. Geophys. Res. {\bf 111} A08106 %DOI: 10.1029/2005JA011303
\end{thebibliography}
\end{document}